\documentclass[a4paper,12pt]{article}

\usepackage{natbib}

\usepackage{pifont}
\usepackage{authblk}
\usepackage{setspace}                                      
\usepackage{amsmath, amsthm, amssymb}
\usepackage{bbm}
\usepackage{braket}
\usepackage{ragged2e}

\usepackage{epigraph}
\usepackage[left=1.2in,right=1.2in,top=1in,bottom=1in]{geometry}                                  
\usepackage{enumitem}

\usepackage{hyperref}

\usepackage{color} 
\definecolor{darkblue}{rgb}{0.0,0.0,0.3}
\hypersetup{colorlinks,breaklinks,linkcolor=darkblue,urlcolor=darkblue,anchorcolor=darkblue,citecolor=darkblue}

\usepackage{ifthen} 
\def\eprinttmp@#1arXiv:#2 [#3]#4@{\ifthenelse{\equal{#3}{}}{\href{http://arxiv.org/abs/#1}{arXiv:#1}}{\href{http://arxiv.org/abs/#2}{arXiv:#2 [#3]}}}
\newcommand{\eprint}[1]{\eprinttmp@#1arXiv: []@}
\newcommand{\doi}[1]{\href{http://dx.doi.org/#1}{doi:#1}}

\def\citepos#1{\citeauthor{#1}'s (\citeyear{#1})} 

\title{{\bf $\Psi$-Epistemic Quantum Cosmology?}}

\author[1]{\bf Peter W. Evans\thanks{email: \href{mailto:p.evans@uq.edu.au}{p.evans@uq.edu.au}}}
\author[2]{\bf Sean Gryb\thanks{email: \href{mailto:s.gryb@hef.ru.nl}{s.gryb@hef.ru.nl}}}
\author[3]{\bf Karim P. Y. Th\'ebault\thanks{email: \href{mailto:karim.thebault@bristol.ac.uk}{karim.thebault@bristol.ac.uk}}}

\affil[1]{\small{{\it School of Historical and Philosophical Inquiry}, University of Queensland}}
\affil[2]{\small{\it Institute for Mathematics, Astrophysics and Particle Physics}, Radboud University}
\affil[3]{\small{{\it Department of Philosophy}, University of Bristol}}

\date{\today\vspace{-3ex}}

\begin{document}

\maketitle

\begin{abstract}
  This paper provides a prospectus for a new way of thinking about the wavefunction of the universe: a $\Psi$-epistemic  quantum cosmology. We present a proposal that, if successfully implemented, would resolve the cosmological measurement problem and simultaneously allow us to think sensibly about probability and evolution in quantum cosmology. Our analysis draws upon recent work on the problem of time in quantum gravity, upon causally-symmetric local hidden variable theories, and upon a dynamical origin for the cosmological arrow of time. Our conclusion weighs the strengths and weaknesses of the approach and points towards paths for future development.
\end{abstract}

\newpage

\setlength{\epigraphwidth}{.5\textwidth}
\epigraph{\justify\textit{Cosmologists, even more than laboratory physicists, must find the usual interpretive rules of quantum mechanics a bit frustrating.}}{J.S Bell first words of `Quantum Mechanics for Cosmologists', 1981}

\setlength{\epigraphwidth}{.5\textwidth}
\epigraph{\justify\textit{Finally, I should mention the semiphilosophical issues arising when one attempts to apply a probabilistic theory to the Universe, of which one has only a single copy. Here I made no attempt to deal with these issues and took a simple-minded approach that the theory describes an ensemble of Universes.}}{A. Vilenkin (a cosmologist) last words of `Interpretation of the wave function of the Universe', 1989}

\tableofcontents

\section{The Proposal}

The interpretational problems that have long plagued the foundations of quantum mechanics are much exacerbated when we attempt to apply quantum theory at a cosmological scale. In particular, if we apply quantum theory to the whole universe, then we are confronted by the three interrelated problems of making sense of measurement, probability, and evolution. This paper provides a prospectus for a new way of thinking about the wavefunction of the universe: a $\Psi$-epistemic  quantum cosmology. We present a proposal that, if successfully implemented, would resolve the cosmological measurement problem and simultaneously allow us to think sensibly about probability and evolution in quantum cosmology.

Our proposal depends upon three distinct interpretational moves. The first move, discussed in \S\ref{sec:first move}, is to implement quantization such that there is evolution in the universal wavefunction. This is in contrast to the more standard `timeless' Wheeler-DeWitt approach to quantum cosmology, but in line with the `Relational Quantization' program developed and defended by \cite{gryb:2011,gryb:2014,Gryb:2015,Gryb:2016a,Gryb:2016b}. The second move, discussed in \S\ref{sec:second move}, is to take an epistemic stance regarding both the universal wavefunction itself and the Born probabilities that are derived on its basis. We propose a $\Psi$-epistemic quantum cosmology in that we hold that the wavefunction of the universe is a bookkeeping device for agential knowledge and not an ontological object. The relevant Born rule probabilities are then also epistemic since they are presumed to be defined in virtue of constraints upon the knowledge of an ideal (classical) epistemic agent embedded within the universe. The viability of this second move is inspired by $\psi$-epistemic approaches to non-relativistic quantum theory. In particular, we will make reference to local hidden variable approaches that exploit the causal symmetry `loophole' in the Bell, Kochen-Specker and PBR no-go theorems.\footnote{This loophole is exploited by denying the implicit assumption of strictly forwards-in-time causality \citep{Costa_de_Beauregard,aharonov:1964,Werbos73,Cramer,Price}. It is distinct from its contrapositive, sometimes called the `free-choice' loophole, which leads to superdeterministic local hidden variables approaches \citep{Bell81,Bell90,Norsen11}. We return to this matter in \S\ref{sec:second move}.} Below we will argue that there are good reasons to believe that an epistemic interpretation of the universal Born probabilities cannot be consistently applied in a quantum cosmology that is timeless. Our first move, therefore, creates conceptual space for the second.

The first two moves lead us to a $\Psi$-epistemic quantum cosmology with both the dynamics of the wavefunction and the projection postulate understood in terms of the changes in the (objective) epistemic constraints placed upon an embedded agent. What, however, is the physical basis behind the existence and change of these epistemic constraints? This is where our third, crucial, and most speculative move is made. In \S\ref{sec:third move}, drawing upon recent work on a \textit{dynamical} origin for the cosmological arrow of time \citep{Barbour:2014}, we suggest that it is plausible to identify the direction of time with the increasing availability of useable records (i.e., complexity) within the context of a causally symmetric local hidden variable model of the universe. As complexity increases, the precision with which we can parametrize our ignorance of the universal hidden variable distribution increases. Thus, as agents, our epistemic situation is linked to the arrow of time and codified in the universal wavefunction. This third and final move resolves a potential worry about the first: that the evolution of the wavefunction of the universe with respect to a time parameter breaks the general covariance of the classical cosmological model. Since the wavefunction is epistemic, it need not be invariant under the same set of symmetries as our ontology -- given by the local hidden variables. Moreover, it might even be taken as a necessary precondition of our states of knowledge that their evolution is defined relative to a simultaneity class. Thus, what might look like a fault in our proposal turns out to be rather an attractive feature.

Together the package of ideas presented in this paper might seem rather too new and ambitious. Our proposal, after all, is to try and solve the measurement problem in quantum cosmology, the problem of time in quantum gravity, and the problem of the cosmic arrow of time -- \textit{simultaneously}. But perhaps these problems have proved so resistant to solution precisely because they have been approached piecemeal. Our main purpose in this paper is to put forward a new interpretational stance and marshal a range of philosophical and physical arguments in its support. At the least, we take ourselves to have established our ambitious proposal to be \textit{not-implausible}. Full assessment requires the construction of a concrete cosmological model. The final section, \S\ref{sec:final}, sets out the conceptual and formal problems that would need to be solved before such a model could be constructed.

\section{Time and Probability in Quantum Cosmology}
\label{sec:first move}
\subsection{The Timeless Probability Problem}

A fundamental principle in all quantum theories is the principle of superposition: the state of the system $\Ket{\psi}$ can consist of a superposition of distinct eigenstates. In a quantum cosmology the quantum state, $\Ket{\Psi}$, is (or at least can be) in a superposition of all the degrees of freedom of the entire universe.\footnote{Here, and throughout the paper we will use $\Psi$ to refer to the quantum state of the entire universe and $\psi$ to refer to the quantum state of a sub-system of the universe.} If we apply standard canonical quantization techniques to general relativity\footnote{Here we are referring specifically to the `Dirac quantization' that is based upon the constrained Hamiltonian theory of constraints \citep{Dirac:1964,Henneaux:1992a}. Application of constraint quantization to the canonical formulation of general relativity leads directly to an equation of the Wheeler-DeWitt form \citep{DeWitt:1967,Thiemann:2007}.} then we get an equation for the quantum state of the whole universe called the Wheeler-DeWitt equation:
\begin{equation}\label{eq:WDE}
  \hat{H}\Ket{\Psi} = 0,
\end{equation}
where $\hat{H}$ is an operator version of the Hamiltonian constraint of canonical general relativity. This equation does not describe any temporal evolution of the quantum state -- it describes a timeless universe. The Wheeler-DeWitt equation gives us a nomological restriction on the state of the universe -- but nothing more. Thus, the problem of interpreting this equation is a fearsome one: it seems very hard to reconcile its structure with our manifestly temporal and non-superposed phenomenology.

One candidate interpretation of the Wheeler-DeWitt equation is the \textit{Many Instants Interpretation} (MII) due to \citet{barbour:1994,Barbour:2003}. The interpretation aims to make sense of superpositions along similar lines to the Everett or many worlds interpretation. The recovery of temporal phenomenology is then tackled by a new idea: `time capsules'. Here we discuss the MII in some detail in order to illustrate the structure of the set of conceptual problems that face the interpretation of timeless quantum cosmology \textit{in general}.

The bare structure of MII is very simple. All we have is: i) a space representing every possible configuration of the universe; and ii) a state functional which is a solution to the Wheeler-DeWitt equation. We then apply the (Born) rule that the modulus squared of the state vector associated with any configuration gives a probability weight to that configuration. The first obvious problem is to make sense of temporal phenomenology with such a sparse, timeless structure. Barbour does this using his time capsules proposal: i) Certain configurations are such that they encode temporal structure -- i.e., records that describe dynamical evolution; ii) The solutions to the Wheeler-DeWitt equation are such that the probabilities (Born rule weights) associated with these time capsule configurations are much higher than those associated with other configurations; iii) The peaking of these probabilities on time capsules is taken to explain our phenomenal experience without the need for time. Explicit calculations by \citet{Halliwell:2001} based upon a mini-superspace model in fact support Barbour's intuition regarding the peaking of amplitudes on time capsules (see also \cite{Anderson:2009,Gomes:2016}). Repetition of this calculation for full quantum cosmology is a difficult challenge.

Despite its appeal, the time capsule has been criticised on a number of conceptual grounds.  In particular, there are worries related to making sense of traditional accounts of personal identity and probability.  Ordinarily we think of observers as enduring embodied things that maintain their identities through time. If we are wrong to think of observers in this way, how can we think of them? An immediate problem with the many instants interpretation of quantum gravity is that it seems not to allow for any conventional notion of personal identity. Such a notion is traditionally conceived of as being \textit{at least partially} dependent upon the continuity of a particular arrangement of matter through time. We cannot have personal identity in anything like this sense in Barbour's framework.\footnote{For critical discussion of this point see in particular \cite{Healey:2002}. We should note there is some hope to deal with this problem via the Saunders-Wallace-Parfit strategy for thinking about Personal identity in Everettian quantum mechanics: `Even in classical physics, it is a commonplace to suppose that transtemporal identity claims, far from being in some sense primitive, supervene on structural and causal relations between momentary regions of spacetime...it is the survival of people who are appropriately (causally/structurally) related to me that is important, not my survival per se' \citep{Wallace:2007}.} The probability problem seems even more acute. As noted succinctly by Dowker:

\begin{quotation}
  The most serious problem is that in a scheme...in which all the possibilities are realized, there is no role for the probabilities. The usual  probabilistic Copenhagen [quantum mechanics] predictions for the results of our observations cannot be recovered...when \textit{all} the elements in a sample space of possibilities are realized, then probability is not involved. [italics in original]\footnote{This text appears in the context of an email exchange with Barbour reprinted in the paperback edition of `The End of Time' \cite[p.355]{Barbour:2003}}
\end{quotation}

There is an obvious challenge to making sense of probabilities in the context of many instants. Clearly these probabilities cannot be understood as long term frequencies since our framework does not allow for any temporal process that selects instants in turn. Furthermore, since \textit{all} instants are actualised there is clear prima facie coherence to Dowker's charge that the use of the word probability is simply not appropriate to this context. We can place the various different notions of probability in the context of two distinctions: epistemic vs ontic and subjective vs objective. Since we are dealing with physical theory we will presume that all the notions of probability we are dealing with are objective so we have two possibilities:
\begin{enumerate}
  \item [A.] Objective Ontic Probability: defined in virtue of the bare ontological structure of the physical theory.
  \item [B.] Objective Epistemic Probability: defined in virtue of the bare ontological structure of the theory combined with constraints placed upon the knowledge of an ideal (classical) epistemic agent embedded in the system described by the theory.
\end{enumerate}

\noindent We will drop the term `objective' for the rest of the discussion and simply refer to these two notions of probability as ontic and epistemic.

One ontic option, due to Barbour himself, is that the Born weights are measures of how many copies of each instant are included in the space of all possible instants. Although there does seem some intuitive attractiveness to the idea of just defining an ontic probability measure in this way, in order to be noncircular we need independent reasons for the space to have this `copies' structure --  and these are currently lacking. Furthermore, as a probabilistic explanation the idea of multiple copies does not make sense without some stochastic process by which one from the array of copies is chosen -- this would seem to require a process \textit{in time} which is not given in the theory. An alternative way of bringing in an ontic notion of probability is to invoke Popper's idea of probability as \textit{propensity} and consider our framework as providing a single case probability corresponding to the propensity of an instant being realised.\footnote{For discussion of the propensity notion of probability in the context of quantum mechanics see \cite{Suarez:2007}.} However, a propensity-type account of universal chances for a timeless universe would have neither explanatory force nor offer any plausible heuristics for theory development. It should thus, on our view, be discounted.

So perhaps we would be better off looking for an understanding of timeless probability at the epistemic level. However, without a solution to the personal identity problem it is not clear that we can genuinely understand these probabilities as referring to `agents'. Moreover, even if we were to coherently define the notion of an agent in the timeless context, it is not clear that such a move would help in the interpretation of the Born rule in the context of many instants. This is because it is precisely the Born rule that must be invoked to assume the perspective of an agent in the first place: there seems to be a fundamental structural problem in giving explanatory priority to epistemic probability in this context. This suggests a general principle defining a limit to the explanatory role of epistemic probability with regard to the probabilistic structure of a physical theory:
\begin{quote}
  \textbf{Primary Precept (PP*)}: an epistemic notion of probability cannot be used to account for ontic probabilistic features of a theory that are themselves necessary to invoke the perspective of agents within the theory.
\end{quote}
If we accept the primary precept\footnote{This is essentially the complement to Lewis' Principal Principle (PP): that (in our terminology) effectively means that epistemic probability should be constrained by ontic probability. See \cite{dawid:2015} for discussion of the primary precept in the context of decoherence and the probability problem in many worlds quantum theory.} then it seems that we are forced to accept that the problem of making sense of probabilities in the context of the many instants interpretation must be tackled at the ontological level.

It is plausible to take this as a general feature of the interpretation of probability in the context of timeless quantum cosmology. A general and necessary requirement of any timeless quantum cosmology is the recovery of the existence of agents with the impression of change. If the mechanism for this recovery involves probabilistic notions then, by \textbf{PP*}, these probabilistic notions must be ontic. We can explicitly reconstruct this argument as follows:
\begin{itemize}
  \item [P1.] A timeless theory of quantum cosmology requires a mechanism to recover the existence of agents with the impression of change \textit{[No Temporal Solipsism]}
  \item [P2.] \textit{(Plausibly)} Any such mechanism will involve probabilistic notions being applied in the context of the wavefunction of the universe.
  \item [P3.] An epistemic notion of probability cannot be used to account for ontic probabilistic features of a theory that are themselves necessary to invoke the perspective of agents within the theory. \textit{[Primary Precept]}
  \item [C.] The reconstruction of the impression of change in timeless quantum cosmology requires a non-epistemic (i.e., ontic) approach to probabilities applied in the context of the wavefunction of the universe.
\end{itemize}

While we do not rule out the possibility that a coherent interpretation of ontic probabilities in the context of the wavefunction of the universe might be possible, we find it difficult to imagine what these `universal chances' might be. Presuming that one wants to avoid this conclusion, and given that both the \textit{No Temporal Solipsism}  and \textit{Primary Precept} premisies must be accepted, this leaves two options. One can \textit{either} attack the relatively weak second premise of the argument, and claim that, in actual fact, we can find a non-probabilistic means of recovering time from within a timeless quantum formalism;\footnote{This option would seem to be in the same spirit as the approach of Valentini (see \cite{Valentini:2005,Towler:2011,Colin:2015,Underwood:2015} and also \cite{Struyve:2010}) within which the Born rule is taken to have a dynamical origin. The extent to which such a framework can be successful implemented in the context of Wheeler-DeWitt cosmologies remains to be seen.} \textit{or} look for a timeful model of quantum cosmology within which epistemic probabilistic notions become plausible. That is, if time is part of the basic ontology of our theory from the start, then we do not need to give arguments (probabilistic or otherwise) as to why agents have the impression of change -- there really is change! In this paper, we will explore the possibilities that lie within this final course, and, in doing so, develop a proposal for a dynamic, $\Psi$-epistemic approach to quantum cosmology.

\subsection{Schr\"{o}dinger Evolution for the Universe}
\label{subsec:pot}

In the previous section, we introduced the Wheeler-DeWitt formalism via a conditional statement: \textit{if} we apply standard canonical quantization techniques to general relativity \textit{then} we derive a timeless equation for the wavefunction of the universe. What if we \textit{do not} apply standard canonical quantization techniques? Are there alternative views on the quantization of general relativity that lead to timeful rather than timeless quantum cosmologies?

One popular option is to move away from the canonical formalism altogether. Rather, many contemporary approaches to quantum gravity are based upon a path-integral type quantization of the covariant formulation of gravity. Approaches along these lines are, for example, causal set theory \citep{bombelli:1987,dowker:2005,henson:2006}, causal dynamical triangulation \citep{loll:2001,ambjorn:2001}, spin foams \citep{baez:1998,perez:2013}, or functional RG approaches \citep{lauscher:2001}. There is much diversity within this broad family of `covariant approaches' to quantum gravity, and evaluation of the extent to which each of them do or do not lead to a genuinely timeful quantum cosmology is a large project that we will not undertake here. We would, however, suggest that those that are genuinely timeful might be amenable to a $\Psi$-epistemic interpretation along the lines of the proposal discussed in the following sections.

Our present project is concerned with interpreting a recently proposed timeful approach to the canonical formalism. This option is largely unexplored despite a number of appealing and attractive features. In particular, given that the standard canonical quantization procedure can be amended such that non-trivial time evolution of the wavefunction remains, we would then have a formalism for quantum cosmology with unitary `Schr\"{o}dinger evolution' of a form analogous to that found within non-relativistic quantum theory. One of the great benefits of retaining such evolution is that a greater number of the candidate interpretations of quantum mechanics could then be extended to the cosmological realm. The particular option that we will explore in the subsequent sections is the `causally symmetric' approach endorsed by \cite{Price} and \cite{Wharton10b}. It is only by implementing Schr\"{o}dinger evolution for the universe that such an interpretation of the universal wavefunction becomes viable.  How then can this move to an amended canonical quantization procedure be justified? Following the work of \cite{gryb:2011,gryb:2014,Gryb:2015,Gryb:2016a,Gryb:2016b} we can put forward an argument towards timeful canonical quantum cosmology based upon three interpretational and formal steps.

The first step draws upon a particular `moderate relationalism' about time. In `radical relationalism' about time we assert that what it means for a physical degree of freedom to change is for it to vary with respect to a second physical degree of freedom; and there is no sense in which this variation can be described in absolute, non-relative terms. This radical relationalism about time is closely associated with the work of \citet{Rovelli:1990,Rovelli:1991,Rovelli:2002,Rovelli:2004,Rovelli:2007,Rovelli:2014}. Formally, we can capture the essence of radical relationalism very concisely in terms of a prescription for the fundamental Hamilton-Jacobi equation. According to Rovelli the hallmark of a relational system of mechanics is that the Hamilton-Jacobi principal functional, $S(t,q,Q)$, can be identified with the characteristic functional  $W(q,Q)$.\footnote{See in particular \cite[\S3.2]{Rovelli:2004}.} That is, rather than solving an equation of the form:
\begin{equation}\label{eq:full HJ}
H\Big{(}q,\frac{\partial S(t,q,Q)}{\partial q}\Big{)}=\frac{\partial S(t,q,Q)}{\partial t}
\end{equation}
via the usual \emph{Ansatz} $S(t,q,Q)=Et+W(q,Q)$, we only have a `timeless' equation of the form:
\begin{equation}\label{eq:red HJ}
H\Big{(}q,\frac{\partial S(q,Q)}{\partial q}\Big{)}=0\,,
\end{equation}
where $S(q,Q)$ in \eqref{eq:red HJ} plays the role of $W(q,Q)$ in \eqref{eq:full HJ}. This radical relationalist way of thinking about the Hamilton-Jacobi formalism  leads naturally to the equations of the Wheeler-DeWitt type, such as \eqref{eq:WDE}, since there is a reliable heuristic, dating back to Schr\"{o}dinger, that takes us from the Hamilton-Jacobi principal functional to the wavefunction \cite[pp. 99-109]{Rund:1966}. Radical relationalists, like Rovelli, embrace a form of timelessness even at the classical level -- for them only relative variation exists, and thus neither the Hamilton-Jacobi principal functional nor the wavefunction of the universe should have any time dependance.

In a moderate relationalist approach, rather than temporal change being based merely on relative variation, such change is argued to be primitive in the sense that it is definable independently for any physical degree of freedom in isolation. This view is defended as a relationalist view on time, rather than a Newtonian absolutist view, on the grounds that although change itself is taken to be primitive, the quantification of change in terms of a temporal measure of duration is still taken to be purely relative. More formally, on the moderate temporal relationalist view defended by Gryb and Th\'{e}bault, there is always assumed to exist a monotonically increasing time parametrization, but this parametrization is taken only to be defined up to diffeomorphism. This view can be reconciled with the formal arguments regarding the Hamilton-Jacobi formalism since the difference between Equations \eqref{eq:full HJ} and \eqref{eq:red HJ} above is entirely due to an extra time boundary term, namely: the shift $S \to S + Et$, which does not affect the local equations of motion. At the classical level the two formalisms are observationally indistinguishable. See \citet{Gryb:2016a} for extensive discussion of this point.

Given the moderate relationalist view on time, one can motivate an alternative canonical quantization procedure for the class of theories that are globally time reparametrization invariant. This is the second formal step in our argument towards timeful quantum cosmology and involves the implementation of a procedure called `relational quantization'. The standard Dirac constraint quantization involves promoting all (first class) canonical constraint functions from the classical theory to operators that annihilate the wavefunction. The rationale behind this is a connection between canonical constraint functions and redundant `gauge' degrees of freedom on phase space. In theories that are globally time reparametrization invariant the Hamiltonian is always a canonical constraint function. There are good reasons, however, to believe that this constraint function has nothing to do with redundant `gauge' degrees of freedom. Rather, the single Hamiltonian constraint of a globally time reparametrization invariant theory results directly from the fact that the relevant Lagrange density is homogeneous of order one in the velocities. By Euler's homogeneous function theorem this implies that the Hamiltonian density must vanish \citep{Dirac:1964}.

In fact, there is no good formal reason to believe that there are redundant degrees of freedom on phase space connected to a global Hamiltonian constraint.\footnote{For extensive formal analysis of this point see \citet[\S3-4]{gryb:2014} and \citet[\S4]{Gryb:2016a}. Also see \cite{Barbour:2008}.} Application of the Dirac quantization procedure to globally time reparametrization invariant theories can only be motivated, if it can be motivated at all, by adoption of radical relationalism about time. Given we adopt moderate relationalism about time we have good reason to look for an alternative quantization strategy under which time is retained. The relational quantization procedure developed by Gryb and Th\'{e}bault is precisely such a strategy. It can be motivated in the context of an analysis of globally time reparametrization theories via Faddeev-Popov path integral \citep{gryb:2011}, constraint quantization \citep{gryb:2014} or Hamilton-Jacobi techniques \citep{Gryb:2016a}. In each case, the resulting quantum formalism retains a fundamental notion of time evolution: unitary evolution of the Schr\"{o}dinger-type,
\begin{equation}
  \hat{H}\Ket{\Psi} = i\hbar\frac{\partial\Ket{\Psi}}{\partial t}.
\end{equation}
This is of course in line with a classical Hamilton-Jacobi formalism understood in terms of Equation \eqref{eq:full HJ} rather than Equation \eqref{eq:red HJ}.

The third and final step is the move that allows this Schr\"{o}dinger-type evolution to be applied to the entire universe. An immediate and obvious limitation in the relational quantization procedure that we have just discussed is that it is designed for \textit{globally} time reparametrization invariant theories with a single Hamiltonian constraint that generates \textit{global} time evolution. General relativity is a \textit{locally} time reparametrization invariant theory with an infinite family of Hamiltonian constraints that generate \textit{local} `many fingered' time evolution. Recent work, however, points towards the viability of reformulating general relativity as a globally reparametrization invariant theory.

The final move is to adopt a re-description of gravity in terms of a formalism that features a notion of preferred slicing. One attractive possibility along these lines is suggested by the \textit{shape dynamics} formalism originally  advocated by Barbour and collaborators (\cite{barbour:scale_inv_particles,Barbour:2011dn,barbour_el_al:scale_inv_gravity,barbour_el_al:physical_dof}) and then brought into modern form in \cite{gryb:shape_dyn}. Within this formalism, the principle of local (spatial) scale invariance is introduced with the consequence of favouring a particular notion of simultaneity. This selects a unique global Hamiltonian and thus allows for relational quantisation to be applied. Shape dynamics is based upon a re-codification of the physical degrees of freedom of general relativity via exploitation of a duality between two sets of symmetries. Whereas general relativity is locally time reparametrization invariant and spatially diffeomorphism invariant; shape dynamics is globally time reparametrization invariant, spatially diffeomorphism invariant, and locally scale (i.e., conformally or Weyl) invariant. In the class of spacetimes where it is possible to move from one formalism to the other (those that are `CMC foliable') the physical degrees of freedom described by the two formalisms are provably equivalent, they are merely clothed in different descriptive redundancy.

Our adoption of the shape dynamics formalism at this stage is not a necessary move -- any formalism for gravity with a preferred notion of simultaneity could be adopted. The unique and attractive feature of shape dynamics is that a preferred slicing is argued for on the basis of a symmetry principle rather than the introduction of preferred observers or other absolute structures. Nevertheless, there is still the worry that, in applying relational quantization to shape dynamics, or any other theory of gravity with a preferred time slicing, we might break the duality between the two sets of symmetries. In particular, it might be worried that the Schr\"{o}dinger-type evolution of the wavefunction of the universe with respect to a time parameter will break the general covariance of the classical cosmological models. It is precisely in response to this concern that the notion of a $\Psi$-epistemic quantum cosmology becomes particularly appealing. But before we can assess this option we must consider the fortunes of $\psi$-epistemic approaches to quantum theory in general.

\section{Local Hidden Variable Quantum Theory}
\label{sec:second move}
\subsection{The Einstein-Bell Conditions}

Consider the following conditions for an interpretation of quantum cosmology, which we call the \textit{Einstein-Bell conditions}:
\begin{enumerate}
  \item No universal chances [\textit{God does not play dice}];
  \item The universe is local;
  \item Quantum cosmology is consistent with the no-go theorems.
\end{enumerate}
The first condition encodes the basic assumption of our research: that, at the level of the whole universe, any probabilistic concepts must be given an epistemic interpretation. We find the concept of universal chances obscure and thus propose to investigate its alternative. The second assumption is a consistency requirement between the quantum cosmological formalism and the general theory of relativity from which we derive our \textit{empirically well-confirmed} classical cosmological models. We take the statement that the universe is local to mean that the ontology of the universe is such that only arrangements of matter and interactions consistent with the causal (e.g., light-cone) structure of Lorentzian spacetimes are permitted. That is, physical influences and bodies (including observers) can only follow time-like or null spacetime trajectories. The third requirement derives from the expectation that any quantum theory of cosmology will display correlations that violate the Bell inequalities. For this reason, we insist that the three basic `no-go' results of contemporary quantum theory -- Bell, Kochen-Specker, and Pusey-Barrett-Rudolph (PBR) -- will all apply in quantum cosmology also. In this section, we will argue that there is a unique realist interpretational stance that satisfies the Einstein-Bell conditions: the causally symmetric hidden variable approach. In order to motivate this conclusion we must first define a taxonomy for interpreting the quantum mechanical wavefunction.

Following \citet{HarriganSpekkens10}, we can distinguish between $\psi$-ontic and $\psi$-epistemic interpretations of the wavefunction. Referring to the complete physical state of some quantum system at some specified time as the `ontic state' of that system, we call the wavefunction description of that system $\psi$-ontic if every distinct quantum state is consistent with a single ontic state. We can further distinguish $\psi$-ontic interpretations into $\psi$-complete and $\psi$-incomplete interpretations. A $\psi$-complete interpretation takes the quantum state to provide a complete description of `reality' (there is a one-to-one correspondence between ontic states and distinct quantum states), while a $\psi$-incomplete interpretation requires that the quantum state be supplemented with additional ontic degrees of freedom. Many of the more well-known interpretations of quantum mechanics are $\psi$-ontic interpretations: many worlds and dynamical collapse interpretations are typically $\psi$-complete, and pilot-wave interpretations are typically $\psi$-incomplete as the `corpuscles' (a.k.a beables) provide additional ontic degrees of freedom over and above the ontic state.

The role of probabilities in $\psi$-ontic interpretations ranges from the straightforward to the obscure.\footnote{See \cite{timpson:2011} for a review.} In the context of dynamical collapse, probability is an inherently  ontic concept: the probabilities are objective chances primitively posited in the theory. Contrastingly, within the pilot-wave interpretation, although the wavefunction is ontic, the probabilities are essentially epistemic: they arise on the basis of $\psi$-incompleteness and reflect our ignorance of the full ontic degrees of freedom. The question of how we are to understand probability in the context of many worlds has been a topic of much vigorous debate that we will not attempt to review in detail here. Some critics of the approach argue along the lines of the Dowker quote above: the concept `probability' is not even an appropriate concept in the context of a quantum formalism where all possibilities are realised \citep{Kent:2010}. Some advocates, on the other hand, claim not only to be able to give a decision theoretic derivation of the Born rule in a many worlds context \citep{Deutsch:1999,Saunders:2004,Wallace:2007}, but also to be able to establish these probabilities as ontic \citep{Wallace:2012}. The basis for Wallace's argument is the close connection between the weights that feature within the Everettian branching structure and the objects that play the functional role of probability for agents.\footnote{In this sense we can see his analysis as at least partially in the same spirit as that of \cite{Vaidman:2012}. We should note, however, that Vaidman emphasises the epistemic rather the ontic aspects of his treatment of probability.} According to Wallace, since branch weights are part of the bare structure of many-words quantum theory, the probabilistic concepts to which they are connected should be taken to be ontic (in our terminology).

In more general terms, since it is a $\psi$-complete interpretation we take it as prima facie reasonable to assume that
\textit{if} there is a probabilistic concept at play with many worlds quantum theory, \textit{then} this concept will be an ontic one. Thus, should one be looking for an epistemic understanding of probabilities in the context of $\psi$-ontic interpretations, we take it that the most plausible option is a pilot-wave type approach.

We call the wavefunction description of some system $\psi$-epistemic when multiple distinct quantum states are consistent with a single ontic state, warranting an interpretation of the wavefunction as a representation of an observer's knowledge, rather than a representation of reality. Since specifying the wavefunction does not completely specify the ontic state, $\psi$-epistemic interpretations are naturally $\psi$-incomplete. We can further distinguish $\psi$-epistemic interpretations into realist interpretations, whereby there exists an underlying ontic state, and anti-realist or operationalist interpretations that make no such claim for a deeper underlying reality. Anti-realist interpretations include (arguably) orthodox Copenhagen interpretations as well as quantum Bayesianism\footnote{Although, \citet{Fuchs16} argues that this is a common misconception of Quantum Bayesianism. Interestingly, the `participatory realism' detailed in \citep{Fuchs16} has notable metaphysical similarities to the causally symmetric local hidden variable theories discussed in this work.} and other quantum informational approaches. Realist interpretations are not well explored. This is despite the fact that, as \citet{HarriganSpekkens10} argue, a realist $\psi$-epistemic interpretation is precisely what Einstein was advocating in his more sophisticated arguments for the incompleteness of quantum mechanics. According to the realist $\psi$-epistemic view, quantum mechanics is a statistical theory over the ontic states which are `hidden' from the observer; the complete theory is then a hidden variable theory. The basic probabilistic concepts that occur in both realist and anti-realist $\psi$-epistemic interpretations are clearly themselves epistemic.\footnote{N.b. there is nothing in principle stopping a realist $\psi$-epistemic interpretation having supplementary stochastic structure within the dynamics of the hidden variables and so having further ontic probabilities.} Thus, as could be expected, $\psi$-epistemic interpretations are natural bedfellows for epistemic notions of probability.

There is perhaps good reason for the lack of exploration of realist $\psi$-epistemic interpretations of quantum mechanics. The development of quantum mechanics was followed from the outset by a series of no-go theorems that seemingly ruled out a range of $\psi$-epistemic hidden variable approaches. The first of these was \citepos{von_Neumann} theorem that the quantum statistics could not arise from an underlying set of determined hidden variables, apparently ruling out hidden variable approaches all together. However, \citepos{Bohm} (albeit $\psi-$ontic) model of quantum mechanics is just such a description of hidden variables that reproduces the quantum statistics, only it is explicitly nonlocal. As a consequence of Bohm's counterexample, a second no-go theorem arises, Bell's theorem \citep{Bell64}, which states that there can be no hidden variable model of quantum mechanics that obeys Bell's notion of local causality, whereby spacelike separated events must be independent conditioned on a past common cause. A further no-go theorem, the Kochen-Specker theorem \citep{Kochen_Specker}, states that a hidden variable model must be contextual, whereby two operationally equivalent experimental preparation procedures may correspond to inequivalent ontic state representations (the state additionally depends on the context of measurement).

A more recent no-go theorem, the PBR theorem \citep{PuseyBarrettRudolph}, states that the ontic states of any interpretation of quantum mechanics that fits within the Bell framework and reproduces the Born rule must be in one-to-one correspondence with the quantum states; that is, the interpretation must be $\psi$-ontic. Given this series of no-go theorems for $\psi$-epistemic hidden variable approaches, it is little wonder they remain under explored. There is, however, one such approach that evades these no-go theorems: causally symmetric local hidden variable approaches to quantum mechanics.\footnote{For discussion of the relation between PBR and causally symmetric  approaches see \cite{Leifer:2011,wharton:2014}.}

Causally symmetric local hidden variable (CSLHV) approaches to quantum mechanics take advantage of a `loophole' in the assumptions that underlie the no-go theorems, assumed most explicitly in Bell's theorem. Not only does Bell assume local causality, he also assumes what could be called the `free variables' assumption \citep{Norsen11} or, equivalently, measurement independence, whereby any hidden variables must remain independent of the choice of measurement settings to which the system is subject as part of the experimental procedure. Relaxing this assumption amounts to allowing the ontic state underlying the quantum description of a system to be directly dependent upon the measurement settings to which it will be subject in the future. But disavowing this assumption can be interpreted ambiguously: a statistical dependence between the ontic state and the measurement settings superficially appears to be suggesting that experimenters are no longer free to choose the measurement settings arbitrarily, resulting in what \citet[p.~244]{Bell90} called `superdeterminism'. Such an interpretation, however, blindly adheres to the implicit assumption of strictly forwards-in-time causality. Another way to view the relaxation of the assumption of independence between the ontic state and the measurement settings -- the direct inverse of superdeterminism -- is explicitly to preserve the free choice of experimenters over the experimental settings but reject the assumption of strictly forwards-in-time causality, allowing a causal influence from future to past to accompany the usual causal influences from past to future. The resulting `causally symmetric' approach circumvents the results of Bell's theorem, and so can be a local hidden variable theory (and in doing so solves any apparent tension with relativity), and thus also circumvents the PBR theorem, and so can be a $\psi$-epistemic interpretation of quantum mechanics. A CSLHV approach also contains an explicit contextuality of the ontic state on the experimental procedure, so fits within the bounds given by the Kochen-Specker theorem.

The loophole in Bell's theorem originates in a suggestion in the 1950s from \citet{Costa_de_Beauregard} (a student of de Broglie) in response to \citepos{EPR} argument that quantum mechanics is incomplete. According to the suggestion, causal influences could propagate as both retarded and advanced waves, in a kind of `zigzag', to avoid the problems posed by apparently nonlocal correlations. Two causally symmetric approaches to quantum mechanics that are more well-known today are the two-state vector formalism developed by \citet{aharonov:1964,aharonov:2014,aharonov:2015}, in which forward evolving and backward evolving state vectors combine to produce the intervening quantum state, and the transactional interpretation developed by \citet{Cramer}, in which quantum particle trajectories emerge from a cycle of retarded and advanced waves (see also \cite{kastner:2012}). A third approach arises from \citepos{Price} foundational philosophical work on causally symmetric quantum theory in addition to \citepos{Wharton10b} more recent formal extension of those foundations to develop an approach to quantum mechanics as a two-time boundary problem.

Causally symmetric local hidden variable interpretations are the most plausible option for a realist interpretation of quantum cosmology satisfying the Einstein-Bell conditions. Epistemic probabilities are not appropriate for a $\Psi$-complete interpretation of the universal wavefunction: any probabilities must be ontic probabilities on such a view. The first Einstein-Bell Condition thus rules out $\Psi$-complete interpretations and restricts us to $\Psi$-incomplete interpretations.\footnote{We note again the subtleties regarding what, if any, interpretation of probability is appropriate in the context of many worlds theory. See \cite{Vaidman:2012,Wallace:2012}.} The combination of the second and third conditions then rules out $\Psi$-ontic interpretations altogether since these cannot be both local and avoid the no-go theorems. This then leaves us with either anti-realist $\Psi$-epistemic interpretations, for example quantum Bayesianism, or local-realist $\Psi$-epistemic interpretations, the only examples of which are causally symmetric local hidden variable approaches (in which the ontic state is taken to be comprised of spatiotemporally local classical variables, in accord with the Einstein-Bell conditions). If one wants to be an Einstein-Bell realist about quantum cosmology, then on our view a CSLHV approach is the most natural way to go.

\subsection{Perspectivalism and the Past}

In the previous section, we motivated the CSLHV approach based upon the combination of realism with the Einstein-Bell conditions for quantum cosmology. In the present section, we will discuss in detail a particular variant of the CSLHV family: the `Price-Wharton' picture. The formal motivation of the Price-Wharton picture is based on Hamilton's principle with emphasis on the constraint of both initial and final boundary conditions to construct equations of motion from a Lagrangian.\footnote{We will return to the analysis of the formal motivation behind the Price-Wharton picture in \S\ref{sec:third move}} If we treat external measurements as physical constraints imposed on a system in the same way that boundary constraints are imposed on the action integral of Hamilton's principle, we can imagine the dynamics of a system subject to preparation and measurement procedures to emerge \emph{en bloc} as the solution to a two-time boundary problem. Focussing solely on classical fields, \citet{Wharton10b} argues that constraining such fields (which characterise the ontic state) at both an initial and a final temporal boundary (or a closed hypersurface in spacetime) generates two strikingly quantum features: quantization of certain field properties and contextuality of the unknown parameters characterising the field between the boundaries. Thus, a classical field constrained at both an initial and a final temporal boundary permits, by construction, ontic variables that are correlated with the future measurement of the system. The final measurement does not simply reveal preexisting values of the parameters, but \emph{constrains} those values (just as the initial boundary condition would) -- thus, had the final measurement been different, the ontic state would have been different, rendering the picture `causally symmetric'.

Within the Price-Wharton picture, an invariant joint probability distribution associated with each possible pair of initial and final conditions can be constructed \citep[p.~318]{Wharton10}, and the usual conditional probabilities can be formed by conditioning on any chosen portion of the boundary \citep[p.~280]{Wharton10b}.\footnote{This interpretation of probabilities maps nicely to the Feynman path integral representation of joint probabilities, wherein the joint probability of particular initial and final state pairs naturally incorporates two temporal boundary conditions, and is given by an integral over the classical action.} As a result, probability is interpreted as a manifestation of our ignorance: if we knew only the initial boundary, we would only be able to describe the subsequent ontic state probabilistically (since we lack knowledge of the final constraint). We thus interpret the solution to the Schr\"{o}dinger equation $\psi$-epistemically as just such a description: it is an ignorance function over the unknown ontic state based on our knowledge of the initial boundary (and lack of knowledge of the final boundary). Once we obtain knowledge of the final boundary, our knowledge of the ontic state undergoes discontinuous Bayesian updating and we can then retrodict the field values between the two boundaries. There is, however, no such discontinuous evolution of the underlying ontic state. Moreover, we could equally conditionalise on the final boundary to generate a probabilistic description propagating backwards in time, but this is rarely useful in practice on account the (assumed) forwards-in-time-facing agential perspective.

It is worth noting at this point (we will return to this issue later) that the Price-Wharton picture forces us to draw a sharp distinction between the determination of behaviour of the quantum state -- the epistemic quantum wavefunction description -- and the underlying ontic state. When we consider determination of the behaviour of the quantum wavefunction description, then since the Schr\"{o}dinger equation is parabolic, specifying the wavefunction solution on an initial boundary is sufficient to specify completely the behaviour of the wavefunction description thereafter. Thus, the Schr\"{o}dinger equation and knowledge of a wavefunction description on a Cauchy surface amount to a well-posed Cauchy problem (and thus the wavefunction description renders quantum mechanics Markovian). One of the lessons of Bell's theorem is that it is not possible according to such a well-posed Cauchy problem for the wavefunction description to be comprised of classical, spatiotemporally located variables -- initial data of this form cannot account for the complete observed quantum behaviour thereafter of any purported classical variables. According to the Price-Wharton picture, however, we take the wavefunction description to be $\psi$-epistemic and, thus, a representation of our knowledge of an underlying ontic state. If we are to think of this ontic state along the lines of Einstein-Bell realism, then it must be the case that specifying the ontic state completely on a Cauchy surface is insufficient for determining the subsequent behaviour of this state; we additionally require information on a future boundary to obtain complete determination. Thus, complete specification of the ontic state on a Cauchy surface combined with whatever dynamical laws govern the ontic state variables cannot amount to a well-posed Cauchy problem. In other words, the laws governing the ontic state variables cannot be parabolic or hyperbolic PDEs.\footnote{This is not strictly the case. It may indeed be possible to maintain the advantages of a causally symmetric approach and still have any subsequent local hidden variables solve a Cauchy problem. The tension here is ultimately between the solution of a Cauchy problem from freely, arbitrarily, and (ideally) completely specifiable initial data and the symmetric expectation that the final boundary be equally freely, arbitrarily and completely specifiable. All that is required to ease this tension is some as yet unidentified constraint on our ability to freely, arbitrarily and completely specify such data -- a not inconceivable possibility.\label{fn:Cauchy}} This issue proves the biggest challenge to construction of a coherent $\Psi$-epistemic quantum cosmology and we will return to it in \S\ref{sec:final}.

It is a fundamental assumption of the Price-Wharton picture that we are ignorant of the future but not the past. The wavefunction is $\psi$-epistemic because it is an ignorance function over the unknown ontic state based on our knowledge of data on some Cauchy surface \textit{and lack of knowledge of data on some future boundary}. The supposed explanation for this asymmetry is grounded in a form of perspectivalism about temporal asymmetry. Perspectivalism about temporal asymmetry is based upon a particular way of combining a `block universe' model of time with an `interventionist' account of causation. According to the former, all past, present and future events are equally real and we imagine time as ontologically on a par with a fourth dimension of space. According to the latter, we say that some event is a cause of some other event when, given an appropriate set of independence conditions, an intervention to manipulate the first event is an effective means of manipulating the second event. This provides the justification for characterising the Price-Wharton picture as causally symmetric. More precisely, $X$ is a cause of $Y$ just in case there is some possible (or hypothetical) intervention $I$ that can be carried out on $X$ that will change the probability distribution over the outcomes at $Y$, so long as $I$ excludes all other possible causes of $X$, $I$ is correlated with $Y$ only through $X$, and $I$ is independent of any other cause of $Y$.

The interventionist account is thus a counterfactual account of causation and is not explicitly reliant on a particular temporal direction to define causation. The direction of causation is dictated by the nature of the functional dependences between the relevant variables describing a system and the nature of the relevant intervention. This permits us to understand causation as a `perspectival' notion, wherein we have a spatiotemporally constrained perspective within the block universe -- we have limited epistemic access to other spatiotemporal regions, especially future regions -- such that when we act as agents there are specific natural constraints on which parts of our environment we take to be fixed and which parts we take to be controllable. It is the epistemic relation that we hold with respect to the different variables involved in the intervention that align the direction of causation with the future temporal direction. We control the intervention and, thus, usually know its significant preconditions. However we \text{do not} have epistemic access to the effect of the the intervention in the future independently of this control.

All together, the Price-Wharton picture provides an attractive interpretational package. It allows us to combine an epistemic interpretation of the wavefunction with a local realist ontology without contravening the no-go theorems. Whilst some \citep[e.g.,][]{Maudlin} have characterised the resulting `retrocausality' as a high ideological cost, and therefore wholly unappealing, \citet{Evans15} points out that in fact there is no ideological cost at all to the Price-Wharton picture of retrocausality (in particular, the view does not countenance `spooky' backwards-in-time effects); rather, this view is simply a natural consequence of our limited epistemic viewpoint within a metaphysically acausal block universe. In the context of cosmology, the block universe view is not just plausible but almost unavoidable.\footnote{Although, see \citet{Petkov07}, especially \citet{Ellis07}, and also \citet{Earman08}, for a discussion of the plausibility of a rival dynamic, or `growing', block universe.}

\section{Constructive Perspectivalism and the Cosmic Arrow}
\label{sec:third move}

Is it viable to apply a causally symmetric local hidden variable interpretation at the cosmological scale? In particular, can we plausibly use the Price-Wharton picture to underpin a `$\Psi$-epistemic' interpretation of the wavefunction of the universe, analogously to $\psi$-epistemic interpretations of the quantum mechanical wavefunction? Whilst we find many aspects of the Price-Wharton picture appealing, we do not think that, as it stands, it is satisfactory for interpreting timeful quantum cosmology. The aim of this section is to explain why, and in doing so lay the foundations for a `constructive extension' of the Price-Wharton picture that we hope this paper to be a first step towards.

One feature of the Price-Wharton picture that strikes us as deeply unsatisfying is the particular notion of perspectivalism that is put forward. In the arguments presented above, a key question was left open: \textit{why} do we have limited access to the future in a different sense to the past? Price's answer, inspired by Boltzmann, is that this asymmetry is a feature of \textit{creatures like us} living in a universe with a particular entropy gradient \cite[pp.278]{Price:2007}:

\begin{quotation}
We regard the past as fixed because we regard it as knowable, at least in principle. This is clearly an idealisation, but one with some basis in our physical constitution. As information-gathering systems, we have epistemic access to things in (what we call) the past; but not, or at least not directly, to things in (what we call) the future.

Plausibly, this fact about our constitution is intimately related to the thermodynamic asymmetry, at least in the sense that such information-gathering structures could not exist at all, in the absence of an entropy gradient. Although the details remain obscure, I think we can be confident that the folk physics reflected in the temporal asymmetry of our epistemic and deliberative templates does originate in \textit{de facto} asymmetries in our own temporal orientation, as physical structures embedded in time.
\end{quotation}

Price's concern in this matter is the asymmetry of causation: why should causes typically precede their effects? His answer is that the asymmetry of causation is deeply rooted in the inherent asymmetry of deliberation, whereby an agent can only deliberate about a desired outcome of some set of possible actions when the actual outcome is unknown to the agent (deliberation is useless where an agent knows the actual outcome in advance). The significant feature of this `architecture of deliberation' is that we regard the past as knowable, and so we typically deliberate towards the unknown future. As the above quote suggests, that we know about the past and not the future is, according to Price, a function of the thermodynamic asymmetry and the entropy gradient; thus, the asymmetry of causation is grounded, via our perspective as temporally embedded agents, in the thermodynamic asymmetry. But the nature of the connection between the thermodynamic asymmetry and our asymmetric epistemic relationship to the past and future has been left largely unexamined.\footnote{For more on Price's views on this issue, see \citet[p.321-5]{Price94}, \citet[p.7-8]{Price07a}, \citet{Price:2007}, \citet[p.199]{Price13} and \citet[p.436-40]{PriceWeslake}.}

There is a suggestive analogy between this move of Price, and Einstein's `self-confessed sin' of treating rods and clocks as `primitive entities' assumed to have a proscribed relationship with the spacetime metric \citep{brown:2005,giovanelli:2014}. This move was an essential part of Einstein's `principled' derivation of special relativity and could be contrasted with a constructive derivation in which the chronometric properties of clocks are derived based upon matter theory, rather than inserted as an assumption. Rather than postulating clocks and rods as primitive entities, in a constructive derivation of relativity theory one would provide a dynamical model of their properties.\footnote{See \citet{bell:1976,miller:2010} for exploration of the idea of a constructive derivation of special relativity.} Similarly, rather than postulating creatures like us with a temporal orientation, one would like to provide a dynamical model in which such creatures in our special epistemic circumstance naturally emerge. In particular, rather than postulating that the cosmic arrow of time, in terms of an entropy gradient, leads to the epistemic constraints that define our temporal orientation, one would wish to give a constructive model of this connection. Although perhaps equally forgivable, Price's sin, in not providing details of this connection, is of a similar magnitude and significance to Einstein's. But can it be corrected?

A necessary requirement for a constructive explanation of temporal orientation at the cosmic scale is the provision of a physical model for the creation of records.  By definition, records are physical structures from which one can retrodict earlier states of the system in which one is embedded. Consider the situation in which there are no records: a universe containing a thermal, completely homogeneous spatiotemporal distribution of the microphysical degrees of freedom. Within such a universe there is no local structure that can be utilised to retrodict past microstate of the universe based upon local observations. There are no records. Now imagine there is some non-trivial structure formation. If, using that structure, an embedded agent can make retrodictions of the past state, then that structure functions as a `record'. It also, due to the time reversal invariance of the underlying equations, will function as a `prophecy' -- a record of the future. Thus, the formation of records in-and-of-itself can tell us nothing about the asymmetry of knowledge since any local record-type structure that forms will always be equally useful for prediction as it is for retrodiction. The creation of records is necessary but not sufficient for constructive temporal perspectivalism.

Rather than the mere creation of records, what we require is an `arrow of complexity' in the records that are created. That is, we need a physical model in which records that are created can be used to make progressively more and more precise retrodictions. They will also, of course, be useful to make more and more precise predictions -- they will function as more and more complex prophecies as well as records. However, the epistemic standpoint of an agent embedded in a universe with such an arrow of complexity will change and this change will have a temporal orientation. As the complexity of the records increases, the `fine-graining' that the agent can make on their own phase space will increase: they will become less ignorant of their own past and future. The point is not that an agent will ever, at a given time, be more ignorant of the orientation we label the past than the orientation we label future. Rather, agents progress between instantaneous temporal states corresponding to increasingly better epistemic standpoints. The time direction of this progression is the arrow of time. In the future, we will be less ignorant about the past.

There is, however, a simple, yet quite devastating, problem afflicting any cosmic model along the lines of that described above. Given that the fundamental equations of cosmology are time reversal invariant, if we use our model to derive an arrow of complexity along one temporal orientation, we can always take a $t\rightarrow-t$ isometry and derive an arrow of complexity along the opposite temporal orientation. Such problems have been much discussed in the context of statistical mechanical `derivations' of the second law of thermodynamics and seem to us insoluble within a standard cosmological framework. In particular, in the context of an open universe with one `past' and one `future', the hope that a time symmetric model can be used to derive an arrow of time is surely a vain one.\footnote{We say this in full awareness of the subtle and sophisticated debate over whether special initial conditions might allow one to derive an arrow of time in the context of a time symmetric universe. From our perspective, since over a century of debate has already been devoted to the issue, without a clear answer, there is an urgent need for a new approach -- hence our desire to look to non-standard cosmological frameworks. See \cite{loschmidt:1876,Price,albert:2001,north:2002,Callender:2004,price:2004,Earman:2006,callender:2010,Wallace:2010} for details of the various relevant disputes -- including questions over the coherence of assigning an entropy to the whole universe in the first place.} The prospects for constructive perspectivalism based upon an `arrow of complexity' thus rest largely upon non-standard cosmological frameworks.

One promising such framework is that provided by the `BKM' model of \cite{Barbour:2014,Barbour:2015sba}. This is a finite dimensional particle model based upon `Machian' relational arguments with regard to both time and scale. This model is thus very closely related to the conceptual foundations of both relational quantization and shape dynamics. In particular, within the BKM particle model, the overall size of the system is not considered a meaningful phase space variable. BKM is thus a \textit{scale invariant particle model} that captures aspects of the limiting behaviour of the shape dynamics description of gravity.\footnote{The important feature of this formalism that leads to the formation of increasingly better records is the existence of attractors on shape space. These attractors result, via Liouville's theorem, from projecting out the scale momentum from the dynamics and are shown to lead to the arrow of complexity.} The first feature of the BKM model that makes it attractive as a basis for constructive perspectivalism is that it effectively defines the Newtonian limit of an open universe with one `past' and \textit{two} `futures'. In this context, one can legitimately appeal to time reversal invariant laws to derive a \textit{dynamical} arrow of complexity, since one is deriving two such arrows, both pointing away from a singular point of minimum complexity. This is precisely what BKM do. In their model, increasingly complex structures form in both temporal directions facing away from a `Janus-point', that plays the role of the big-bang (see, for example, figures 1 and 2 of \cite{Barbour:2014}). In the BKM model, the existence of the Janus point is not assumed as a special initial condition. Rather, in the model, it is a distinguished phase space point which can be proven to exist along generic solutions of the system, and, thus, it is a natural place to define generic initial conditions. While it remains to be seen whether or not this particular model can be extended towards quantum cosmology,\footnote{ The most important obstructions to this are the classical singularity theorems in GR. These can be bypassed either by quantum effects or classically by removing scale (see for example \cite{throughbigbang:2016}).} there are good formal reasons to expect that the `one past, two futures' structure will also be found within relational quantized mini-superspace models for FLRW quantum cosmology -- see \cite{Gryb:2016b}.

The crucial conceptual connection that our interpretation of the BKM model relies upon is between an arrow of complexity and an increase in precision of `fine-graining' that an agent can make of their own phase space based upon the available records. The particular structure within the BKM model that we take to plausibly play the role of records are the Kepler pairs. These are locally bound and approximately stable two-body sub-systems of an N-body system. Our interpretation of Kepler pairs as records (prophecies) that can be used for retrodiction (prediction) relies on treating them as playing the role of both local `rods' and local `clocks'. A clock is defined by the periodic motion of a sufficiently isolated Kepler pair. A rod is the size of the pair itself. The BKM model gives a concrete realisation of `rods' and `clocks' getting `better' with time since the two arrows of complexity are simultaneously aligned with the tightening of the orbit of the pair and the increase in stability of the periodic motion. Along each of the arrows of time in the BKM model, the periodic motion of a typical Kepler pair becomes ever closer to a genuinely isolated system. This means a typical Kepler pair will increasingly march in step with the other pairs. They move more and more effectively as free inertial systems.

The key point for our purposes is that as these `rods' and `clocks' become `better' in this sense, they enable an agent to use them to distinguish more and more phase space solutions. This already gives us the basic structure discussed above: an arrow of complexity aligned to an increasingly better epistemic standpoint for local agents. In the context of this model, the arrow of complexity in fact leads to an even more powerful epistemic temporal asymmetry. This is because the N-body system is generically chaotic, meaning that solutions diverge in phase space exponentially. This exponential divergence implies that the number of solutions the local agent must distinguish between grows faster in the direction of increasing complexity. Thus, although the agent's rods and clocks become better towards the future, the predictive value of any given resolution in the fine-graining of the phase space diminishes. Due to the arrow of complexity, this effect is temporally asymmetric. Looking towards the direction of reducing complexity, the solutions are still spreading out exponentially, however, our ability to distinguish between these solutions is getting worse. As we move away from the Janus point, we have better knowledge of the past solutions compared with that of the future ones. Within this model, the arrow of complexity picks out an arrow of increase in precision of retrodiction, that is \textit{not} symmetric with respect to prediction.

This particular strategy for implementation of constructive perspectivalism relies upon an arrow of complexity combined with the chaotic structure of N-body particle dynamics. Clearly, whether or not such structure can be reconstructed in a relativistic and field theoretic cosmological model remains to be seen. Nevertheless, we take it that our discussion of the BKM model is enough to show that there is a good ongoing basis to pursue concrete models to underpin constructive perspectivalism of the type needed for $\Psi$-epistemic quantum cosmology. This notwithstanding, while the BKM model gives some of the structure that would lead to constructive perspectivalism, it has one key feature that rules it out as a basis for ongoing work: the model solves a Cauchy problem and not a two-time boundary problem. Thus, it is not a viable candidate for the cosmic local hidden variables of $\Psi$-epistemic quantum cosmology. This points to a more general worry of finding consistent models that solve two-time boundary problems. This worry is the major focus of the next section.

\section{$\Psi$-Epistemic Quantum Cosmology?}
\label{sec:final}

In the last three sections, we have developed and defended a novel and, we hope, plausible proposal for $\Psi$-epistemic quantum cosmology. In this section, we will isolate and assess  a number of important challenges to our package of ideas.

The first issue is, in a sense, the most obvious one. In \S\ref{subsec:pot}, we argued in favour of an approach to quantum cosmology in which there is fundamental time evolution in the wavefunction of the universe. Our current best theory of classical cosmology is general relativity within which the fundamental symmetry of general covariance implies that time is `many fingered' -- in particular, the local time reparametrization invariance of the theory    implies that a global time evolution parameter will not in general be well defined. Does it make sense to have a symmetry at the level of the classical theory that is broken in this sense within the quantum formalism? We can now offer a good response to this worry in light of the context of the Price-Wharton picture of non-relativistic $\psi$-epistemic quantum theory.

In the Price-Wharton approach to quantum theory, the wavefunction may evolve non-unitarily upon measurement. Formally speaking, the unitarity of the quantum evolution has deep (and rather complicated) connection to the conditions on the flow of the Hamiltonian vector fields that guarantee consistent classical dynamics \cite[\S5]{landsman:2007}. Thus, there might seem to be a  tension between a classical local hidden variable model with consistent dynamics and a quantum formalism that includes non-unitary evolution. From a $\psi$-epistemic perspective, such a combination is, however, not as problematic as it may seem. If the wavefunction is not something in the world then clearly conditions on the consistency of classical evolution need not be reflected in symmetries of the wavefunction. Non-unitarity relates to discrete changes in an agent's state of knowledge and should not be taken as having implications for the underlying classical dynamics. The Price-Wharton picture allows us to understand the wavefunction as evolving non-unitarily without there being any corresponding incompleteness or inconsistency in the corresponding hidden variable dynamics.\footnote{See \cite{aharonov:2014} for consideration of this issue within the perspective of the two-state vector formalism.} In a similar vein, in $\Psi$-epistemic quantum cosmology, it is entirely consistent to insist that the local hidden variables are generally covariant even while the wavefunction of the universe picks out a preferred cosmological time. Just as the non-unitarity of the evolution of the wavefunction is a function of agential perspective so is, we argue, the existence of a preferred evolution parameter.\footnote{We should note that the issue at hand is a subtle one. In particular, whilst non-unitary quantum evolution does not directly contradict a classical symmetry principle, a preferred time in the context of quantum cosmology certainly would. Thus, the coherence of our timeful $\Psi$-epistemic quantum cosmology rests on a subtle reinterpretation of symmetries in the context of classical-quantum limits.} It might even be taken as a necessary precondition of our states of knowledge that their evolution is defined relative to a simultaneity class. Thus, our proposal for $\Psi$-epistemic quantum cosmology offers a promise of what might be a full resolution of the problem of time in quantum gravity: a coherent conceptual framework for reconciling quantum evolution with a generally covariant classical formalism.

This discussion leads us to a second potential worry; this one much more difficult to deal with. In taking $\Psi$-epistemic quantum cosmology seriously we must reconsider exactly what quantum-classical limiting procedures mean. One of the requirements of the Price-Wharton picture was that the local hidden variables do not obey dynamical equations with a well-posed Cauchy problem.\footnote{Modulo the caveat in fn.~\ref{fn:Cauchy}.} Such a requirement is clearly also necessary in $\Psi$-epistemic quantum cosmology: without it our approach would fall foul of the relevant no-go theorems (i.e., Bell, Kochen-Specker, PBR) that must reasonably be assumed to apply in the cosmological context. However, one would also expect that any sensible hidden variable theory of quantum cosmology must still contain a limit where it is approximately reproducing classical Lorentzian field theory -- for example, electromagnetism. The problem here is that, in this limit, the two-time boundary problem for Lorentzian field theories is not well defined since the field equations are typically hyperbolic due to the Lorentzian signature of the spacetime. There is a worrying tension between the demands that a $\Psi$-epistemic cosmology must \textit{both} describe an underlying dynamics of hidden variables that \textit{do not} solve a Cauchy problem \textit{and} contain a limit where it recovers classical field theories that \textit{do} solve a Cauchy problem. There is, however, a viable route of escape from this seemingly fatal impasse. There are at least two formal resources found within modern physics that one can draw upon to obtain a Lorentzian field theory described by hyperbolic equations from a Euclidean field theory described by elliptic equations. Such resources give us a means to `square the circle' and describe a system as both solving a Cauchy problem (in some limit) but not-solving a Cauchy problem (in the fundamental dynamics). What we have in mind here is using techniques developed in the context of `emergent gravity' and `Wick rotation'. We will spend some time explaining the potential applicability of each approach below.

Within the `emergent gravity' approach it is argued that classical field theories, including electromagnetism and general relativity, can be understood as low energy limits of a field theory of fundamentally different character. Particularly prominent implementations of such an idea include various forms of the `entropic gravity' proposal \citep{jacobson:1995,verlinde:2011}. More straightforwardly, but in the same spirit, one can make the simple observation that if our classical field theories are emergent, in the sense of resulting from structurally different underlying dynamics, then the character of the partial differential equations of Maxwell's theory (for example) might also be an emergent feature. That is, one might imagine that a dynamics described by hyperbolic PDEs might emerge from a more fundamental theory that features PDEs which are elliptic. Such an idea is explicitly examined by \citet{barcelo:2007}, who points out that there is a very large set of systems which can be appropriately described in an averaged fashion by a hyperbolic system of PDEs, even though the fundamental equations are elliptic. The essence of the idea comes from techniques used in the context of analogue gravity,\footnote{The original proposal of analogue gravity was by \citet{unruh:1981}. Reviews are \cite{barcelo:2005,visser:2007}. See \cite{Dardashti:2015a} for a philosophical discussion.} wherein linearised fluctuations in a medium can, under certain conditions, obey effective equations with Lorentzian signature,  while the bulk medium is governed by the (Euclidean) equations of non-relativistic continuum mechanics. In such a context, it is quite plausible for the universe to be described by local hidden variables that do not solve a Cauchy problem, whilst simultaneously there is an emergent classical dynamics that does.

In analogue gravity, a Lorentzian field theory with hyperbolic equations may be understood as \textit{emergent} from an underlying Euclidean field theory with elliptic equations. A much more standard technique for moving between two such systems of equations is `Wick rotation'. Wick rotation is used in quantum field theory to convert a complex Lorentzian path integral to a real Euclidean partition function. In that context, it is a technique used to prove convergence and to control certain divergences of the Lorentzian path integral -- see, for example, \cite{ticciati:1999}. Our claim  is that, on top of being an important tool for rigorously analysing Lorentzian quantum field theories, Wick rotation may also provide a second approach for resolving the apparent tension within our proposal. This is because a Euclidean partition function: i) can be well defined as a two-`time' boundary problem; and ii) can, under certain conditions, be analytically continued to an equivalent Lorentzian path integral.\footnote{More precisely, there are a set of necessary and sufficient conditions under which Euclidean Green's functions are guaranteed to define a unique Wightman quantum field theory. See \cite{osterwalder:1973}.} Moreover, the manner in which the analytic continuation and subsequent complex rotation is performed is intimately connected to the form of the propagators one uses for quantisation. Since these, in turn, directly determine the form of microcausality implemented in the Lorentzian field theory, Wick rotation allows us to get direct access to influences in the Euclidean field theory that could be analytically continued back into causally symmetric influences in the Lorentzian framework. Euclidean field theory has, therefore, features that mark it out as a good starting point to construct a causally symmetric local hidden variable theory.\footnote{Of course, our Euclidean theory should not be fully equivalent to the Lorentzian field theory, else we would end up running into the no-go theorems again. Rather, the idea would be to formulate a theory that although it strictly violates the Osterwalder-Schrader conditions, admits a coarse-grained description in which the partition function can be Wick rotated to a valid Lorentzian field theory.} The fundamental equations of such a theory would be taken to fail to solve a Cauchy problem and yet, after some coarse-graining, lead to a partition function that could be suitably Wick rotated to a Lorentzian path integral.

It is not only the elliptic form of the equations that make the Euclidean formalism attractive from a $\Psi$-epistemic perspective. Unlike its Lorentzian cousin, the Euclidean path integral can be interpreted as a genuine statistical mechanical partition function. That is, one can interpret each path in the sum over histories as a genuine element of a statistical mechanical ensemble since each term of the sum is real and, therefore, there is no interference between individual paths. Additionally, because a coarse-graining\footnote{Or some limiting procedure that would effectively integrate out the retrocausal modes.} would be required to transform a non-Cauchy theory in the Euclidean setting to one that can be analytically continued to a Lorentzian theory, there is a natural way to identify our ignorance in terms of the coarse-grained degrees of freedom. In this context, a statistical interpretation of the resulting coarse-grained theory is perfectly natural. Thus, Euclidean field theory naturally complements a $\Psi$-epistemic interpretation of a local hidden variables theory. Whether motivated by emergent gravity or Wick rotation, in the context of Euclidean field theory the hope of reconciling our set of seemingly irreconcilable desiderata should no longer be taken to be an entirely vain one.

\section*{Final Thoughts}

This paper has included a rather large number of new and controversial ideas. The last being perhaps the most speculative. We feel, however, that we have presented a plausible conceptual platform upon which a $\Psi$-epistemic quantum cosmology might be built. The next step is the construction of a concrete cosmological model which contains all the necessary mathematical and conceptual structures. This will be the focus of future work.

\section*{Acknowledgement}

We are very appreciative of comments on a draft manuscript that we received from Eliahu Cohen, Casey McCoy and Ken Wharton. We are also very grateful for the support of the University of Bristol Institute for Advanced Studies and the School of Arts for supporting this project via the provision of visiting fellowships for two of the authors. S.G. would like to acknowledge support from the Netherlands Organisation for Scientific Research (NWO) (Project No. 620.01.784) and Radboud University. PWE would like to acknowledge support from the Templeton World Charity Foundation (TWCF 0064/AB38) and the University of Queensland.

\bibliographystyle{timebib}
\bibliography{RQCBib}

\end{document}